\documentclass[draft]{agujournal2019}
\usepackage{url}
\usepackage{lineno}
\usepackage{amsmath}
\usepackage{multirow}
\usepackage{amssymb}
\usepackage[inline]{trackchanges}
\usepackage{soul}

\renewcommand{\linenumbers}{}

\draftfalse

\journalname{JGR: Solid Earth}

\makeatletter
\def\ps@headings{\def\@oddfoot{\centerline{\small --\the\c@page--}}%
  \let\@evenfoot\@oddfoot\def\@oddhead{}\let\@evenhead\@oddhead}
\makeatother

\begin{document}

\title{Is the Number of Foreshocks Underestimated? The Role of Declustering Symmetry}

\authors{Herbian Hokky Prabowo\affil{1,2}, Eugenio Lippiello\affil{3}, Sebastian Hainzl\affil{4}, Giuseppe Petrillo\affil{5}}

\affiliation{1}{Asian School of the Environment, Nanyang Technological University, Singapore}
\affiliation{2}{Earth Observatory of Singapore, Nanyang Technological University, Singapore}
\affiliation{3}{Department of Mathematics and Physics, University of Campania “L. Vanvitelli”, Caserta, Italy}
\affiliation{4}{GFZ German Research Centre for Geosciences, Potsdam, Germany}
\affiliation{5}{CNRS, ENS de Lyon, LPENSL, UMR5672, 69342, Lyon cedex 07, France}

\correspondingauthor{Eugenio Lippiello}{eugenio.lippiello@unicampania.it}

\begin{keypoints}
\item A time-reversed nearest-neighbour metric provides a symmetric framework for identifying earthquake foreshocks.
\item Synthetic ETAS and ETAFS tests show rNN reduces foreshock underestimation, especially for small earthquakes.
\item Southern California data show rNN identifies more foreshocks, highlighting declustering symmetry effects. 
\end{keypoints}

\begin{abstract}

The physical origin of foreshocks remains a major unresolved problem in statistical seismology. In bottom-up (BU) triggering models, such as the Epidemic-Type Aftershock Sequence (ETAS) model, foreshocks are ordinary earthquakes stochastically cascading into a larger event. In contrast, top-down (TD) models view them as manifestations of a preparatory process preceding the mainshock. This ambiguity heavily impacts earthquake forecasting and catalog declustering, since inferred foreshock numbers depend on the assumed mechanism.

Here we investigate foreshock detectability using synthetic catalogs from two limiting physical scenarios: purely bottom-up triggering (ETAS) and mixed preparatory-triggering dynamics, implemented via the Epidemic-Type Aftershock and Foreshock Sequence (ETAFS) model. We compare the classical nearest-neighbor (NN) declustering metric with a newly introduced reverse nearest-neighbor (rNN) metric. While NN scales the space-time correlation metric to the magnitude of the earlier event, rNN scales it to the later event. Mathematically, rNN is equivalent to applying the standard NN metric to a time-reversed catalog, establishing a direct symmetry between the two approaches.

Our simulations reveal that declustering performance depends strongly on the magnitude range considered. In both ETAS and ETAFS catalogs, the rNN metric improves the reconstruction of foreshock populations associated with small foreshocks and small mainshocks, reducing the negative bias observed with the standard NN approach. For larger magnitude classes, NN and rNN produce increasingly similar results. These findings suggest that foreshock detectability depends heavily on declustering symmetry and magnitude regime, and that standard methods may underestimate foreshock activity particularly in the low-magnitude regime.

\end{abstract}

\section*{Plain Language Summary}
Foreshocks are earthquakes that occur before a larger earthquake, but identifying them depends on how earthquake catalogues are analysed. Most existing methods assume that earthquakes are linked only forward in time, which may cause some foreshocks to be missed. In this study, we introduce a new approach that applies the same declustering method in reverse time, creating a temporally symmetric alternative called reverse nearest neighbour. Using synthetic earthquake catalogues and observations from Southern California, we show that this method identifies more foreshocks, particularly among smaller earthquakes, while leaving the overall clustering structure largely unchanged. Our results suggest that the reported abundance of foreshocks is not only controlled by earthquake physics but also by the assumptions built into the declustering method, with important implications for earthquake forecasting and for understanding earthquake preparation.

\section{Introduction}

Foreshocks have long occupied a central position in earthquake science because of their potential implications for understanding earthquake nucleation and improving short-term seismic forecasting. Since the early observations of anomalous seismic activity preceding large earthquakes, numerous studies have attempted to determine whether foreshocks constitute a distinct physical phenomenon or simply represent a statistical fluctuation of ordinary seismicity. Successful retrospective examples, such as the 1975 Haicheng earthquake, further strengthened the idea that foreshocks may contain information about impending large ruptures and therefore about the preparatory phase of earthquakes \cite{jones1982foreshock,wang2006predicting}. Yet, despite decades of investigation, the physical nature of foreshocks remains unresolved. \\
Two idealized conceptual frameworks are commonly used to interpret foreshock activity. The first is the bottom-up (BU), or cascade, triggering hypothesis, according to which foreshocks emerge naturally from earthquake-to-earthquake interactions through stress transfer. In this framework, earthquakes are not fundamentally distinguishable as foreshocks, mainshocks, or aftershocks; these labels simply reflect their position within a triggering cascade. Statistical models such as the Epidemic-Type Aftershock Sequence (ETAS) model provide the canonical mathematical realization of this idea \cite{ogata1988,ogata1998}, where each earthquake can trigger offspring events according to scale-invariant temporal, spatial, and magnitude-dependent kernels. Under this interpretation, foreshocks are therefore not fundamentally different from ordinary seismicity and do not necessarily imply the existence of a preparatory process before large earthquakes. This view is also consistent with self-organized criticality approaches to seismicity \cite{bak1988self,bak:1989} and has been supported by several statistical analyses of earthquake catalogs \cite{helmstetter2003,felzer:2004,marzocchi2011statistics}. \\
The second framework is the top-down (TD), or preparatory, hypothesis, which attributes foreshocks to processes associated with the approach to a larger instability, such as progressive stress loading, aseismic slip, slow nucleation, fluid migration, or damage localization \cite{das1981theory,ohnaka1992earthquake,dodge1995foreshock}. In this view, foreshocks are not merely the consequence of stochastic earthquake triggering but may represent localized stress release generated by the evolving state of the future rupture region. Under the TD interpretation, foreshocks may therefore carry genuine predictive information about impending failure. \\
However, this distinction should not be interpreted too rigidly. Recent reviews emphasize that both cascade and pre-slip models represent simplified end-member descriptions of earthquake preparation and that natural sequences often involve multiple interacting processes acting over different spatial and temporal scales \cite{martinez2024cascade}. Stress transfer, aseismic slip, fluid migration, fault heterogeneity, and damage localization may all contribute simultaneously to the preparation of large earthquakes. Consequently, real foreshock sequences need not be purely bottom-up or purely top-down. In the present work, ETAS and ETAFS should therefore be viewed as controlled limiting cases that allow us to investigate a more specific question: how does the detectability of foreshocks depend on the temporal symmetry implicitly assumed by the declustering method? \\
Distinguishing between these competing hypotheses has proven extraordinarily difficult because both approaches can reproduce many observed statistical properties of seismicity. Several studies concluded that ETAS-like triggering adequately explains many observed foreshock sequences and argued that foreshocks may simply be indistinguishable from normal clustered seismicity \cite{helmstetter2003,felzer:2004,hardebeck2008improved,marzocchi2011statistics}. Other studies instead reported deviations from ETAS expectations, including anomalous acceleration patterns, spatial organization, or excess foreshock occurrence preceding large earthquakes \cite{brodsky11,lippiello2012spatial,SZMNW18,PL21,PL23}. Consequently, the question of whether foreshocks represent a genuinely distinct phenomenon remains open. \\
Importantly, this debate has implications extending beyond earthquake nucleation itself. In practice, identifying foreshocks requires a declustering procedure capable of separating background events, aftershocks, and candidate foreshocks. However, declustering methods are not model-independent. Every declustering strategy implicitly assumes a particular physical picture of seismicity. Nearest-neighbor methods \cite{ZB13,ZGKW08}, stochastic declustering approaches \cite{ZOV02}, and ETAS-based classifications are all fundamentally rooted in assumptions about triggering organization in space, time, and magnitude. As a result, if the assumed declustering framework does not match the true physical mechanism generating seismicity, systematic biases may arise in the inferred foreshock population. \\
A related issue was already highlighted by \citeA{christophersen2008foreshock}, who showed that inferred foreshock rates depend strongly on the adopted definitions of foreshocks, aftershocks, and earthquake clusters. Using a consistent cluster-based framework, they demonstrated that the observed foreshock occurrence in Southern California is largely compatible with the expectations of triggering models such as ETAS. Their results suggest that apparent discrepancies between observed and predicted foreshock rates may partly arise from methodological choices rather than from fundamental physical differences. The present study extends this idea by investigating whether the declustering metric itself can introduce additional biases in the inferred foreshock population. \\
This issue is particularly important because most previous studies focused on determining whether foreshocks exist or whether they differ statistically from ordinary seismicity, while comparatively less attention has been devoted to understanding whether the declustering procedure itself may distort foreshock statistics depending on the underlying physical model. In other words, the apparent abundance or scarcity of foreshocks may partly reflect methodological biases rather than intrinsic properties of the earthquake process. \\

In this work, we investigate precisely this problem. Using extensive synthetic simulations, we compare the performance of nearest-neighbor declustering approaches under two limiting physical regimes: a purely bottom-up triggering regime and a mixed preparatory-triggering regime \cite{PL21,PL23}. The corresponding statistical models are introduced in Section Models. This allows us to test whether the inferred number of foreshocks depends on the temporal symmetry assumed by the declustering metric. \\
Our results reveal that the largest differences between NN and rNN are concentrated in the low-magnitude portion of the foreshock-mainshock matrix. In both limiting scenarios, standard NN declustering tends to underestimate foreshocks associated with small foreshocks and small mainshocks, whereas rNN substantially reduces this bias. For larger magnitude classes, the two methods produce increasingly similar results. These findings indicate that conventional declustering approaches may preferentially miss low-magnitude foreshock activity. \\
To address this limitation, we introduce a revised nearest-neighbor metric, denoted rNN (reverse nearest-neighbour), designed to test the sensitivity of foreshock identification to the temporal orientation of the correlation metric. Through systematic numerical experiments, we show that rNN significantly reduces the declustering bias and more accurately reproduces the true foreshock statistics across a broad range of magnitude differences and triggering conditions. \\
As a final step, we apply the same comparison to the Southern California catalogue \cite{HYS12}. In real catalogues, the true genealogy of earthquakes is unknown, and therefore the goal is not to determine which declustering method is definitively correct. Rather, we use Southern California as a practical test case to quantify how strongly the inferred number of foreshocks depends on the adopted declustering symmetry. This application shows that rNN identifies systematically more foreshocks than NN in several magnitude classes, suggesting that standard forward-in-time declustering may underestimate foreshock activity in real seismicity as well.
More generally, our results suggest that foreshock detectability depends on the temporal symmetry of the declustering method and on the magnitude regime considered.

\section{Models and declustering methods}

Let us consider an earthquake catalogue
\[
\mathcal{C}=\{(t_i,\mathbf{x}_i,m_i)\}_{i=1}^{N},
\]
where \(t_i\) is the occurrence time, \(\mathbf{x}_i\in \mathbb{R}^2\) is the epicentral position, and \(m_i\geq m_0\) is the magnitude of event \(i\). Events are ordered in time, so that \(t_i<t_j\) for \(i<j\). In the following, we first introduce the two generative models used to represent the limiting physical scenarios discussed above: ETAS for purely bottom-up triggering and ETAFS for mixed preparatory-triggering dynamics. We then describe the two declustering metrics used to reconstruct foreshock-mainshock relations.

\subsection{The ETAS model}

The Epidemic-Type Aftershock Sequence (ETAS) model is the standard statistical description of earthquake clustering in which seismicity is generated by the superposition of a background process and earthquake-to-earthquake triggering \cite{ogata1988,ogata1998}. In this framework, each earthquake can trigger subsequent events, and the distinction between foreshocks, mainshocks and aftershocks is not intrinsic to the model but emerges only a posteriori from the relative magnitude and time ordering of events.

The conditional intensity for the occurrence of an earthquake of magnitude \(m\geq m_0\), at position \(\mathbf{x}\), and time \(t\), conditional on the history
\[
\mathcal{H}_t=\{(t_i,\mathbf{x}_i,m_i):t_i<t\},
\]
is written as
\[
\lambda_{\rm ETAS}(t,\mathbf{x},m|\mathcal{H}_t)
=
\left[
\mu(\mathbf{x})
+
\sum_{i:t_i<t}
\phi(t-t_i,\mathbf{x}-\mathbf{x}_i;m_i)
\right]
f(m),
\]
where \(\mu(\mathbf{x})\) is the background seismicity rate, \(\phi\) is the triggering kernel, and \(f(m)\) is the Gutenberg--Richter magnitude density. We assume
\[
f(m)=\beta e^{-\beta(m-m_0)},
\qquad
\beta=b\ln 10,
\]
or, equivalently,
\[
f(m)=b\ln(10)\,10^{-b(m-m_0)}.
\]

The triggering kernel is factorized into productivity, temporal decay and spatial decay:
\[
\phi(t-t_i,\mathbf{x}-\mathbf{x}_i;m_i)
=
K_0\,10^{\alpha(m_i-m_0)}
\,g(t-t_i)\,h(\mathbf{x}-\mathbf{x}_i;m_i),
\]
with
\[
g(\tau)
=
\frac{p-1}{c}
\left(1+\frac{\tau}{c}\right)^{-p},
\qquad \tau>0,
\]
and
\[
h(\mathbf{r};m_i)
=
\frac{q-1}{\pi d\,10^{\gamma(m_i-m_0)}}
\left(
1+
\frac{\|\mathbf{r}\|^2}{d\,10^{\gamma(m_i-m_0)}}
\right)^{-q}.
\]
Thus,
\[
\phi(\tau,\mathbf{r};m_i)
=
K_0\,10^{\alpha(m_i-m_0)}
\frac{p-1}{c}
\left(1+\frac{\tau}{c}\right)^{-p}
\frac{q-1}{\pi d\,10^{\gamma(m_i-m_0)}}
\left(
1+
\frac{\|\mathbf{r}\|^2}{d\,10^{\gamma(m_i-m_0)}}
\right)^{-q}.
\]

This formulation corresponds to a purely bottom-up triggering mechanism. Every event may generate descendants, and a foreshock is simply an event whose triggered cascade happens to include a larger earthquake. Therefore, in ETAS, the occurrence of a foreshock--mainshock pair does not require a preparatory physical process. It is instead a rare realization of the same stochastic triggering mechanism that also generates aftershocks.

\begin{table}[htbp]
\centering
\caption{ETAS and ETAFS parameters used in the synthetic simulations.}
\label{tab:etas_parameters}
\begin{tabular}{lc}
\hline
Parameter & Value \\
\hline
$m_0$ & 3 \\
$b$ & 1.06 \\
$\alpha$ & 0.8 \\
$K_0$ & 0.038 \\
$c$ & 0.024 \\
$p$ & 1.27 \\
$d$ & 0.006 \\
$q$ & 1.506 \\
$\gamma$ & 0.85 \\
$K_f$ & 0.05 \\
$\alpha_f$ & 0.54 \\
\hline
\end{tabular}
\end{table}

\subsection{The ETAFS model}

The Epidemic-Type Aftershock and Foreshock Sequence (ETAFS) model extends ETAS by incorporating the possibility that earthquakes are preceded by a preparatory phase. We emphasize that ETAFS is not intended to represent the full complexity of earthquake preparation in nature. Rather, it provides a controlled statistical benchmark in which part of the foreshock population is explicitly organized by the future mainshock. This idealization is useful for testing whether a reverse-time declustering symmetry can recover precursor-like organization while remaining consistent with the view that natural earthquake preparation may involve several interacting mechanisms rather than a single universal process \cite{martinez2024cascade}. The motivation is that, under a top-down loading scenario, foreshocks are not merely ordinary triggering events but are manifestations of an underlying process that culminates in the mainshock. In this view, the final earthquake of a preparatory cluster is not caused by the foreshocks in the same way that an aftershock is caused by a previous earthquake; rather, both foreshocks and the final event are different manifestations of the same evolving instability.

Following \citeA{PL21}, the ETAS intensity is interpreted as the occurrence rate of the final event of a cluster rather than of an isolated earthquake. Each such final event \(k\), with coordinates \((t_k,\mathbf{x}_k,m_k)\), may be preceded by a set of foreshocks distributed in space and time around the final event. The total intensity can therefore be written as
\[
\lambda_{\rm ETAFS}(t,\mathbf{x},m|\mathcal{H}_t)
=
\lambda_{\rm ETAS}(t,\mathbf{x},m|\mathcal{H}_t)
+
f(m)
\sum_{k:t<t_k}
\psi_f(t_k-t,\mathbf{x}-\mathbf{x}_k;m_k).
\]

The foreshock kernel is chosen in analogy with the ETAS aftershock kernel:
\[
\psi_f(\tau,\mathbf{r};m_k)
=
K_f\,10^{\alpha_f(m_k-m_0)}
\,g_f(\tau)\,h_f(\mathbf{r};m_k),
\qquad \tau=t_k-t>0,
\]
where
\[
g_f(\tau)
=
\frac{p_f-1}{c_f}
\left(1+\frac{\tau}{c_f}\right)^{-p_f},
\]
and
\[
h_f(\mathbf{r};m_k)
=
\frac{q_f-1}{\pi d_f\,10^{\gamma_f(m_k-m_0)}}
\left(
1+
\frac{\|\mathbf{r}\|^2}{d_f\,10^{\gamma_f(m_k-m_0)}}
\right)^{-q_f}.
\]
In the simplest implementation, one sets
\[
p_f=p,\qquad c_f=c,\qquad q_f=q,\qquad d_f=d,\qquad \gamma_f=\gamma,
\]
so that the foreshock kernel has the same functional form as the aftershock kernel, but with a different productivity \(K_f\) and magnitude scaling \(\alpha_f\).

Operationally, for each final ETAS event \(k\), the number of foreshocks is drawn from a Poisson distribution,
\[
N_f^{(k)}\sim {\rm Poisson}
\left(K_f\,10^{\alpha_f(m_k-m_0)}\right).
\]
The occurrence times and positions of these foreshocks are then sampled from \(g_f\) and \(h_f\), respectively. Their magnitudes are sampled from the Gutenberg--Richter distribution, with the additional constraint
\[
m<m_k,
\]
so that foreshocks are smaller than the final event of the preparatory cluster.

Although the numerical construction appears to condition present events on a future earthquake, this does not represent a violation of causality. The correct interpretation is that the cluster is the elementary object of the point process. The final earthquake and its foreshocks are different events belonging to the same physical preparation process. Equivalently, one may write the ETAFS model as a marked cluster point process in which an event generated by the ETAS backbone carries an internal structure:
\[
\mathcal{K}_k=
\left\{
(t_{k,1},\mathbf{x}_{k,1},m_{k,1}),
\dots,
(t_{k,n_k-1},\mathbf{x}_{k,n_k-1},m_{k,n_k-1}),
(t_k,\mathbf{x}_k,m_k)
\right\},
\]
with
\[
t_{k,a}<t_k,
\qquad
m_{k,a}<m_k,
\qquad
a=1,\dots,n_k-1.
\]
In this representation, no event is caused by the future. Rather, the whole cluster is generated as one marked object whose last event is the mainshock-like rupture. The top-down nature of the model enters through the fact that the spatial and temporal organization of the foreshocks is controlled by the properties of the final event.

\subsection{Nearest-neighbor declustering}

We next define the declustering procedure used to reconstruct clusters from a catalogue. The nearest-neighbor approach introduced by \citeA{BP04} and further developed by \citeA{ZGKW08,ZB13} assigns to each earthquake a most correlated predecessor. The method is based on the idea that the expected number of earthquakes occurring in a space-time-magnitude domain provides a natural measure of correlation between events.

For two events \(i\) and \(j\), with \(t_i<t_j\), let
\[
\tau_{ij}=t_j-t_i,
\qquad
r_{ij}=\|\mathbf{x}_j-\mathbf{x}_i\|.
\]
The expected number of events with magnitude comparable to \(m_i\), occurring within a time interval \(\tau_{ij}\) and a spatial region of radius \(r_{ij}\), scales as
\[
n_{ij}
\propto
\tau_{ij}\,r_{ij}^{D}\,10^{-b m_i},
\]
where \(D\) is the fractal dimension of epicentres and \(b\) is the Gutenberg--Richter exponent. Up to a multiplicative constant, this defines the nearest-neighbor distance
\[
\eta_{ij}^{\rm NN}
=
\tau_{ij}\,r_{ij}^{D}\,10^{-b m_i},
\qquad t_i<t_j.
\]
Small values of \(\eta_{ij}^{\rm NN}\) indicate that event \(j\) is unusually close in space and time to event \(i\), given the magnitude of \(i\), and is therefore likely to be correlated with it.

For each event \(j\), the nearest predecessor is defined as
\[
i^\star(j)
=
\arg\min_{i:t_i<t_j}
\eta_{ij}^{\rm NN}.
\]
If
\[
\eta_{i^\star(j),j}^{\rm NN}<\eta_c,
\]
then event \(j\) is linked to \(i^\star(j)\); otherwise, \(j\) is considered an independent background event. The directed links \(i^\star(j)\to j\) define a forest of seismic clusters. Within each connected component, the largest earthquake is identified as the mainshock. Events occurring after the mainshock are classified as aftershocks, whereas events occurring before the mainshock and belonging to the same cluster are classified as foreshocks.

This metric is intrinsically oriented forward in time: correlations are evaluated by asking which previous event is the most likely parent of the present event. The magnitude factor \(10^{-b m_i}\) weights the productivity of the candidate parent. Therefore, larger previous earthquakes can collect later events across larger space-time domains. This is physically well suited to aftershock identification, where the causal direction coincides with the temporal direction.

\subsection{Reverse nearest-neighbor declustering}

The standard nearest-neighbor metric is asymmetric in time because it uses the magnitude of the earlier event. This is natural for aftershocks, but it is not necessarily optimal for foreshocks. In a top-down loading scenario, foreshocks are organized by the future mainshock or, more precisely, by the preparatory process that culminates in the mainshock. Therefore, the relevant magnitude scale controlling the spatial and temporal extent of the foreshock cloud may be the magnitude of the later event rather than that of the earlier event.

To capture this reverse organization, we introduce a reverse nearest-neighbor metric, denoted rNN. For two events \(i\) and \(j\), with \(t_i<t_j\), we define
\[
\eta_{ij}^{\rm rNN}
=
\tau_{ij}\,r_{ij}^{D}\,10^{-b m_j}.
\]
The only formal difference with the standard nearest-neighbor metric is that the magnitude entering the Gutenberg--Richter factor is now the magnitude of the later event \(j\), rather than that of the earlier event \(i\).

This construction can be understood as a time-reversed version of the standard nearest-neighbor metric. Let \(\mathcal{T}\) be the time-reversal operator acting on a catalogue as
\[
\mathcal{T}:
(t_i,\mathbf{x}_i,m_i)
\mapsto
(-t_i,\mathbf{x}_i,m_i).
\]
Applying the standard nearest-neighbor metric to the time-reversed catalogue amounts to asking, for each event in reversed time, which previous event is its closest ancestor. In the original time direction, this corresponds to asking which future event best explains the organization of a preceding event. Since under \(\mathcal{T}\) the temporal order of \(i\) and \(j\) is exchanged, the standard metric in reversed time becomes
\[
\eta_{\mathcal{T}j,\mathcal{T}i}^{\rm NN}
=
|t_j-t_i|\,r_{ij}^{D}\,10^{-b m_j},
\]
which is exactly
\[
\eta_{ij}^{\rm rNN}.
\]
Therefore,
\[
\eta_{ij}^{\rm rNN}
=
\eta_{\mathcal{T}j,\mathcal{T}i}^{\rm NN}.
\]

The rNN metric is thus not an ad hoc modification, but the time-reversed counterpart of the standard nearest-neighbor construction. It preserves the same space-time-magnitude scaling as the Baiesi--Paczuski--Zaliapin metric, but reverses the event whose magnitude defines the characteristic correlation volume. In physical terms, NN tests whether a past earthquake is sufficiently large and close to explain a later event as its offspring. By contrast, rNN tests whether a future earthquake is sufficiently large and close to organize previous events as its foreshocks.

For each event \(i\), one may therefore define the most correlated future event as
\[
j^\star(i)
=
\arg\min_{j:t_j>t_i}
\eta_{ij}^{\rm rNN}.
\]
If
\[
\eta_{i,j^\star(i)}^{\rm rNN}<\eta_c^{\rm r},
\]
then event \(i\) is linked to the future event \(j^\star(i)\). Equivalently, this is obtained by applying the standard NN declustering to the time-reversed catalogue and then mapping the links back to the original time orientation.

The resulting directed links point from earlier events toward later events, but their interpretation is different from standard triggering links. They should not be read as causal triggering relations from foreshock to mainshock. Instead, they identify events that are geometrically and statistically compatible with belonging to the preparatory domain of a later earthquake. In this sense, rNN is designed to be sensitive to top-down foreshock organization while retaining the same scaling principles underlying the classical nearest-neighbor method.

\subsection{Comparison between NN and rNN}

The distinction between NN and rNN can be summarized by the two correlation distances
\[
\eta_{ij}^{\rm NN}
=
(t_j-t_i)\,r_{ij}^{D}\,10^{-b m_i},
\qquad t_i<t_j,
\]
and
\[
\eta_{ij}^{\rm rNN}
=
(t_j-t_i)\,r_{ij}^{D}\,10^{-b m_j},
\qquad t_i<t_j.
\]
The standard NN metric assigns the correlation volume to the earlier event and is therefore naturally adapted to bottom-up triggering. The rNN metric assigns the correlation volume to the later event and is therefore naturally adapted to top-down organization.

This distinction is particularly important when the magnitude difference
\[
\Delta m=m_j-m_i
\]
between a candidate foreshock \(i\) and a candidate mainshock \(j\) is large. The two metrics are related by
\[
\eta_{ij}^{\rm rNN}
=
\eta_{ij}^{\rm NN}\,10^{-b(m_j-m_i)}
=
\eta_{ij}^{\rm NN}\,10^{-b\Delta m}.
\]
Therefore, if \(m_j>m_i\), the rNN distance is smaller than the NN distance. This means that rNN naturally enlarges the correlation domain of a future large event when searching backward for its possible foreshocks. Conversely, if \(m_i\) is comparable to \(m_j\), the two metrics become similar:
\[
m_i\simeq m_j
\quad \Rightarrow \quad
\eta_{ij}^{\rm rNN}\simeq \eta_{ij}^{\rm NN}.
\]
This explains why the difference between the two declustering schemes is expected to depend strongly on the foreshock--mainshock magnitude contrast.

In this work, we use synthetic catalogues generated by ETAS and ETAFS to test how these two metrics perform when the underlying physical mechanism is known. Since ETAS catalogues encode bottom-up triggering, they provide a controlled benchmark for the standard NN metric. Since ETAFS catalogues encode a top-down preparatory component, they provide a controlled benchmark for rNN. The comparison allows us to quantify whether the inferred number of foreshocks is a robust property of the catalogue or instead depends on the assumed symmetry of the declustering metric. \\

\subsection{Automatic threshold selection}

In the original nearest-neighbor declustering framework, the metric
\(\eta_{ij}^{\rm NN}\) provides a natural ordering of possible event-to-event
links, but a threshold \(\eta_c\) is still required to decide whether an event
is classified as triggered or background. Traditionally, this threshold has
often been chosen by visual inspection of the bimodal nearest-neighbor
distribution, or by fitting mixture models to the distribution of nearest-neighbor
distances \cite{ZB13}. However, such choices may introduce an additional
source of subjectivity and may affect the inferred number of clusters and,
therefore, the number of identified foreshocks.

In this work, we avoid manually tuning the threshold and use the automatic
non-parametric procedure introduced by \citeA{bountzis2026automatic}. This method is
based on the Susceptibility Index (SI), which quantifies the sensitivity of the cluster structure to variations of the nearest-neighbor threshold. The key idea is to examine how the number of background events changes as progressively
weaker links are added to the catalogue. If the threshold is too strict, many events are classified as independent background events. If it is too loose,
unrelated clusters are artificially merged. The optimal threshold is identified
as the value for which the connectivity structure is least sensitive to adding
new links, corresponding to a minimum of the susceptibility curve.

More formally, let
\[
s_{ij}=\frac{1}{\eta_{ij}}
\]
be the similarity associated with the nearest-neighbor distance. For a given
threshold \(s_{\rm th}\), one constructs a binary similarity matrix by retaining
only links satisfying
\[
s_{ij}>s_{\rm th}.
\]
For each value of the threshold, the number of empty rows of this matrix gives
the number of events that are not linked to any previous event, and are
therefore classified as background events. Denoting this number by \(K(w)\),
where \(w\) parametrizes the threshold level, and denoting by \(N_0(w)\) the
number of retained links, one defines
\[
z(w)
=
-\frac{dK(w)}{dN_0(w)}.
\]
This quantity measures how rapidly the number of background events decreases
as additional links are introduced. The Susceptibility Index is then defined as
\[
\zeta(w)
=
z(w)\,y(w),
\qquad
y(w)=\frac{N_0(w)}{N},
\]
where \(N\) is the total number of events in the catalogue. The optimal
partition is obtained from the non-trivial minimum of \(\zeta(w)\), after
applying the smoothing and shifting procedure described by \citeA{bountzis2026automatic}.

This automatic procedure has two advantages for the present study. First, it
removes the need to choose \(\eta_c\) manually, ensuring that NN and rNN are
compared using the same objective threshold-selection criterion. Second, it
reduces the bias that can arise from arbitrary threshold choices. Tests on large
ensembles of synthetic ETAS catalogues show that the SI method accurately
recovers the true number of background events and achieves classification
metrics above \(90\%\) \cite{bountzis2026automatic}. Therefore, the differences reported
below between NN and rNN should be interpreted as consequences of the
different temporal symmetry of the metrics, rather than of different manual
choices of the declustering threshold.

In the following, the automatic SI procedure is applied separately to the NN
and rNN distances. For NN, the similarity is defined as
\[
s_{ij}^{\rm NN}=\frac{1}{\eta_{ij}^{\rm NN}},
\]
whereas for rNN we use
\[
s_{ij}^{\rm rNN}=\frac{1}{\eta_{ij}^{\rm rNN}}.
\]
The same algorithm is then used to determine the corresponding thresholds
\(\eta_c^{\rm NN}\) and \(\eta_c^{\rm rNN}\).

\begin{figure}[htbp]
\centering
\includegraphics[width=0.48\textwidth]{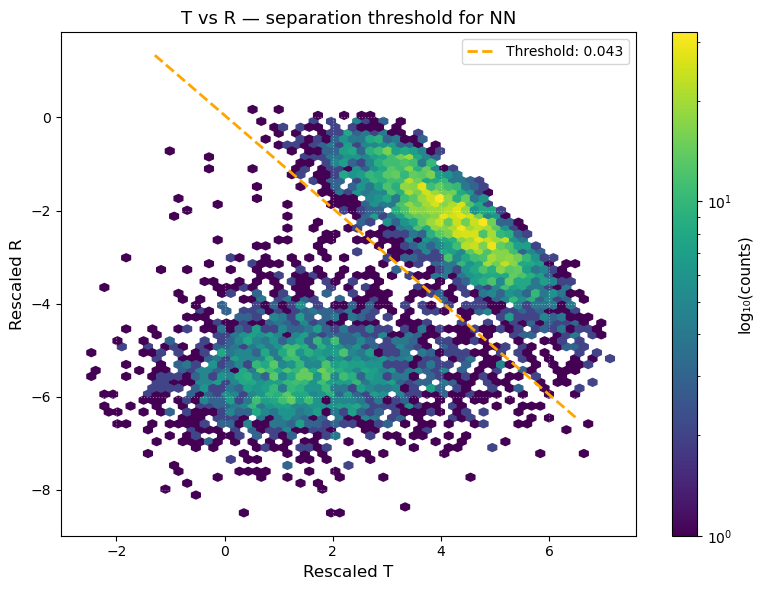}
\includegraphics[width=0.48\textwidth]{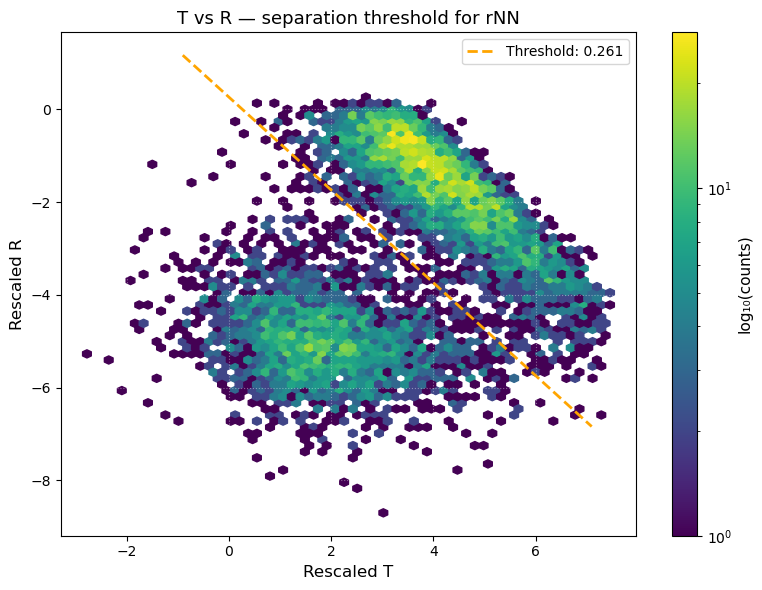}
\caption{
Distribution of nearest-neighbour distances in the rescaled
space-time representation for an ETAS catalogue.
Left panel: standard NN metric.
Right panel: reverse nearest-neighbour (rNN) metric.
The variables \(T\) and \(R\) denote the logarithms of the
rescaled temporal and spatial components of the nearest-neighbour distance, respectively. The dashed orange line corresponds to the automatic separation threshold obtained from the susceptibility-index procedure. Events below the line are classified as triggered,
whereas events above the line are classified as background events. Both NN and rNN display the characteristic bimodal structure previously reported for nearest-neighbour declustering, indicating a
robust separation between clustered and background seismicity.}
\label{fig:TR_NN_rNN}
\end{figure}

Figure~\ref{fig:TR_NN_rNN} shows the distribution of the rescaled temporal and spatial components used in the NN and rNN declustering procedures for an ETAS catalogue simulated with the parameters provided in Table~\ref{tab:etas_parameters}. In both cases, the data exhibit the characteristic bimodal structure reported by
\citeA{ZGKW08,ZB13}, corresponding to clustered and background
seismicity. The automatic SI procedure identifies a separation
threshold that closely follows the valley between the two
populations. Remarkably, the resulting thresholds are very similar for NN and rNN, suggesting that reversing the temporal symmetry of the metric does not substantially modify the global background-triggered partition of the catalogue.

\section{Results}

\subsection{Synthetic catalogue declustering}

We first evaluate the performance of NN and rNN declustering on synthetic catalogues, where the true genealogy of each event is known by construction. This allows us to compare the number of foreshocks inferred by each declustering method with the true number of foreshocks generated by the model. We consider two classes of synthetic catalogues: ETAS catalogues, in which all foreshocks originate from bottom-up triggering, and ETAFS catalogues, in which foreshock activity may include both bottom-up triggering and top-down preparatory clustering. \\
The ETAS and ETAFS synthetic catalogues were generated using the parameter set provided here in Table~\ref{tab:etas_parameters}. These parameters were previously calibrated and validated against Southern California seismicity and were kept fixed throughout this study. Since our objective is to compare the performance of the NN and rNN declustering procedures under identical synthetic conditions, we do not perform a new calibration of the ETAS or ETAFS models.

For each model, we generate \(N_{\rm cat}=1000\) independent synthetic catalogues with the same magnitude threshold \(m_0=3\). In each catalogue, we identify mainshocks and their associated foreshocks using three different definitions. First, we use the true genealogy of the simulation, which provides the reference foreshock set. Second, we reconstruct clusters using the standard nearest-neighbor metric, obtaining the NN-inferred foreshock set. Third, we repeat the declustering using the reverse nearest-neighbor metric, obtaining the rNN-inferred foreshock set.

To quantify the dependence on magnitude, we group mainshocks and foreshocks into discrete magnitude classes. Mainshocks are divided into the classes
\[
M \in [3,4),\ [4,5),\ [5,6),\ [6,\infty),
\]
whereas foreshocks are divided into the classes
\[
F \in [3,3.5),\ [3.5,4),\ [4,4.5),\ [4.5,\infty).
\]
For each pair of magnitude classes \((i,j)\), where \(i\) labels the foreshock magnitude class and \(j\) the mainshock magnitude class, we compare the number of foreshocks inferred by each declustering method with the true number obtained from the synthetic genealogy. For each catalogue and for each method \(k\in\{\mathrm{NN},\mathrm{rNN}\}\), we define the normalized foreshock-count residual
\[
\epsilon_k^{ij}
=
\frac{
N_k^{ij}-N_{\rm true}^{ij}
}{
N_k^{ij}+N_{\rm true}^{ij}
},
\]
where \(N_{\rm true}^{ij}\) is the true number of foreshocks in the magnitude bin \((i,j)\), and \(N_k^{ij}\) is the number reconstructed by method \(k\). Therefore, \(\epsilon_k^{ij}=0\) corresponds to perfect agreement with the true genealogy, \(\epsilon_k^{ij}<0\) indicates underestimation, and \(\epsilon_k^{ij}>0\) indicates overestimation. This normalization avoids the asymmetry and instability of the conventional relative error based solely on \(N_{\rm true}\), particularly for sparse magnitude bins where the true number of foreshocks can be very small.

For each magnitude bin, we then compare the empirical distributions of \(\epsilon_{\rm NN}^{ij}\) and \(\epsilon_{\rm rNN}^{ij}\). A declustering method is considered more accurate when its residual distribution is more sharply concentrated around zero and has smaller systematic bias.

This synthetic setting is essential because it allows us to separate two effects that are inseparable in real catalogues: the physical mechanism generating foreshocks and the ability of the declustering method to detect them. In ETAS catalogues, the true foreshocks are generated entirely through bottom-up triggering cascades. Therefore, any discrepancy reflects a limitation of the declustering method in recovering bottom-up foreshock chains. In ETAFS catalogues, by contrast, the true foreshock population contains events associated with a preparatory top-down component. Therefore, discrepancies in this case quantify how well NN and rNN recover foreshocks generated by a mixed bottom-up/top-down process.

\subsubsection{Stability of the background-triggered partition}

Before comparing the foreshock statistics obtained with NN and rNN declustering,
it is important to verify whether the two metrics produce substantially different
background-triggered partitions of the catalogue. Indeed, a possible concern is
that the differences observed in the inferred number of foreshocks could simply
arise from a radically different classification of earthquakes into background
and triggered events.

\begin{figure}[htbp]
\centering
\includegraphics[width=0.48\textwidth]{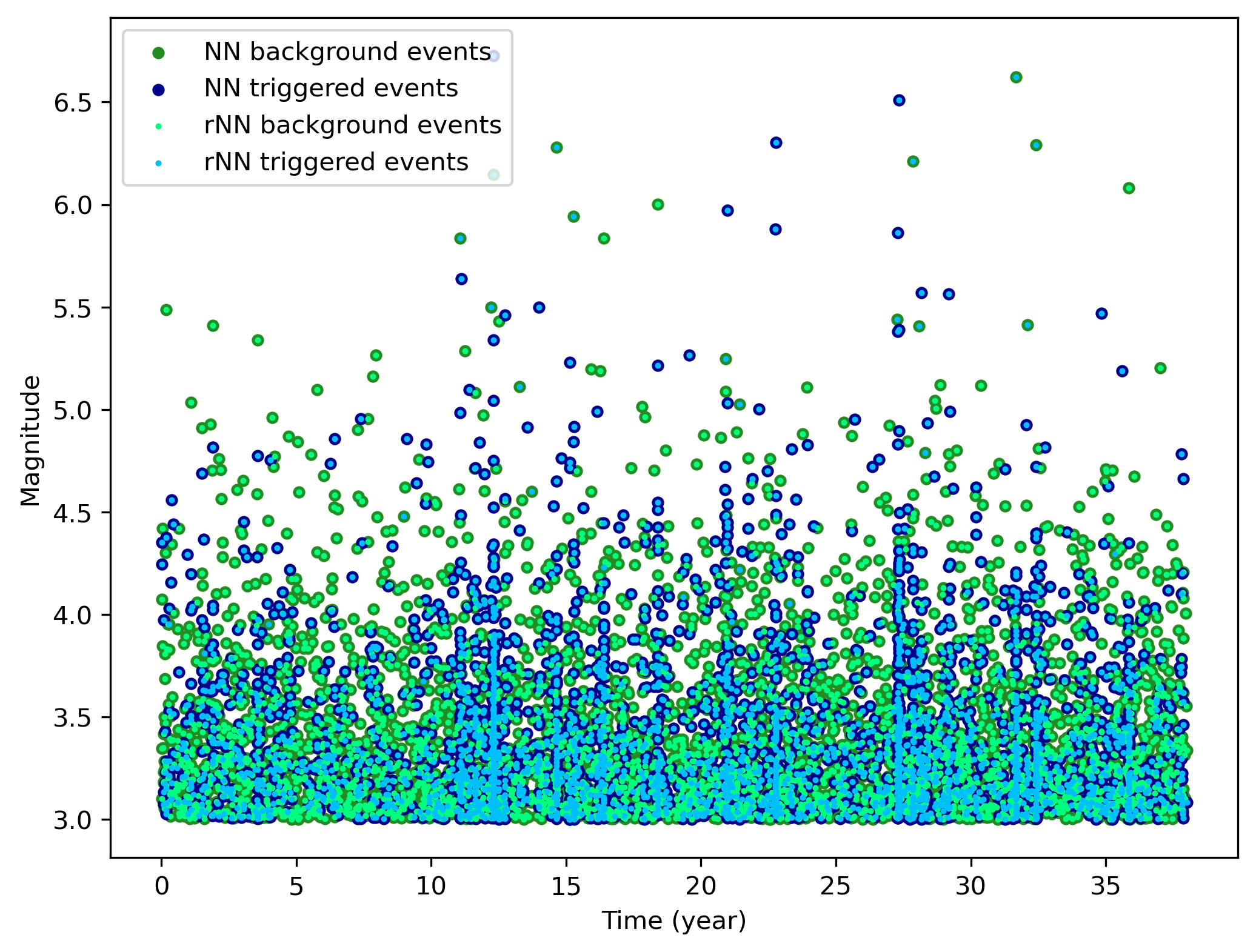}
\includegraphics[width=0.48\textwidth]{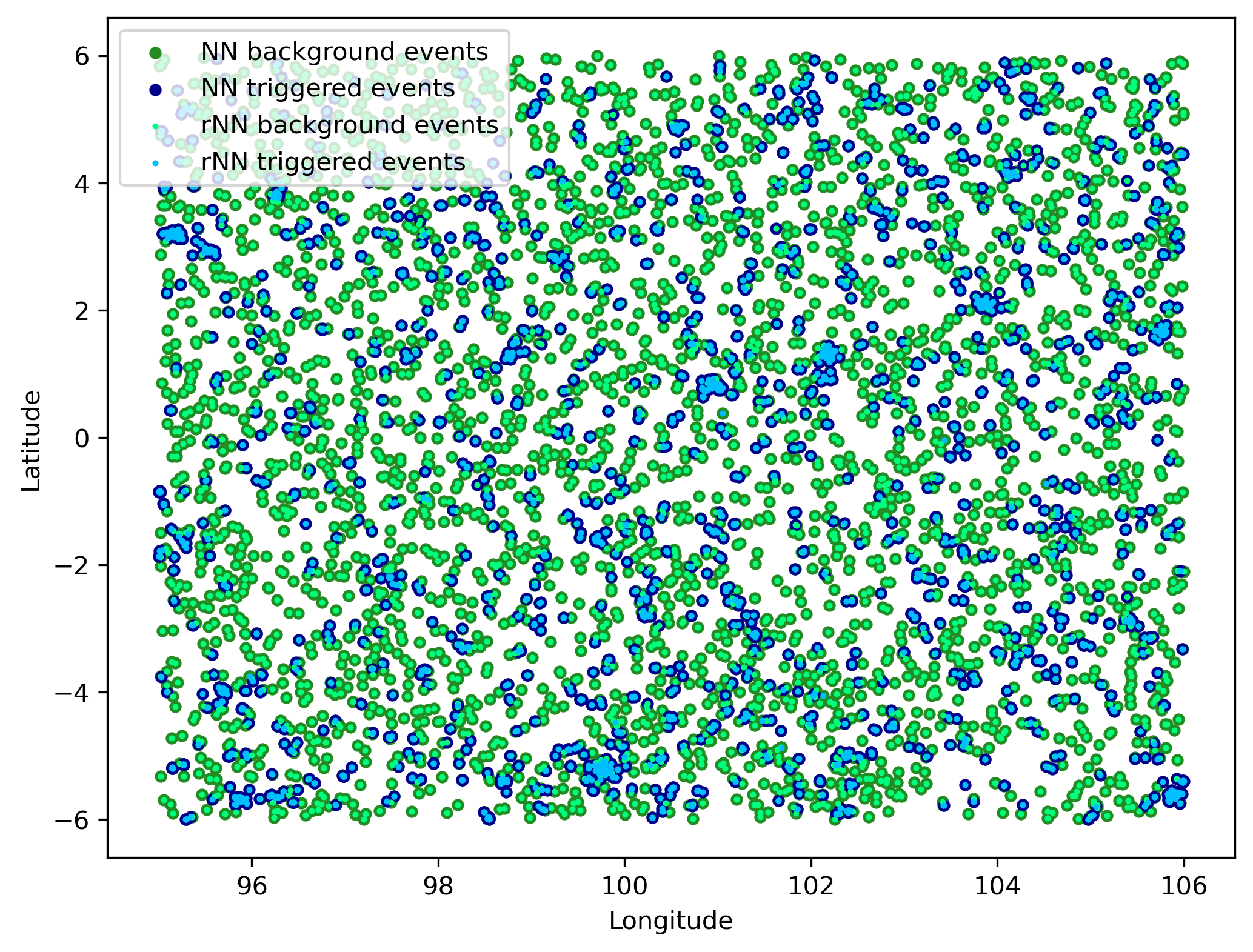}
\caption{
Comparison between the background-triggered partition obtained using the
standard nearest-neighbor (NN) declustering and the reverse nearest-neighbor
(rNN) declustering on the same synthetic ETAS realization.
Left panel: time--magnitude distribution of events.
Right panel: spatial distribution of events.
Big green and small cyan symbols correspond to events classified as background (BG) by NN
and rNN, respectively, while big blue and small light blue symbols correspond to events
classified as triggered (TRIG). The two methods produce nearly identical
background-triggered partitions, with only minor local differences.
}
\label{fig:NN_rNN_partition}
\end{figure}

To test this possibility, we applied both NN and rNN declustering to the same
synthetic ETAS realization and compared the resulting classification of
background and triggered events. Figure~\ref{fig:NN_rNN_partition} shows the
time--magnitude distribution and the spatial distribution of the events
classified by the two methods.

The two declustering methods produce remarkably similar background-triggered
partitions. Both the temporal distribution and the spatial organization of
background events remain nearly unchanged when replacing NN with rNN. Only
minor local differences are observed, mainly close to the declustering
threshold where events can switch classification between background and
triggered. \\

\begin{figure}[htbp]
\centering
\includegraphics[width=0.48\textwidth]{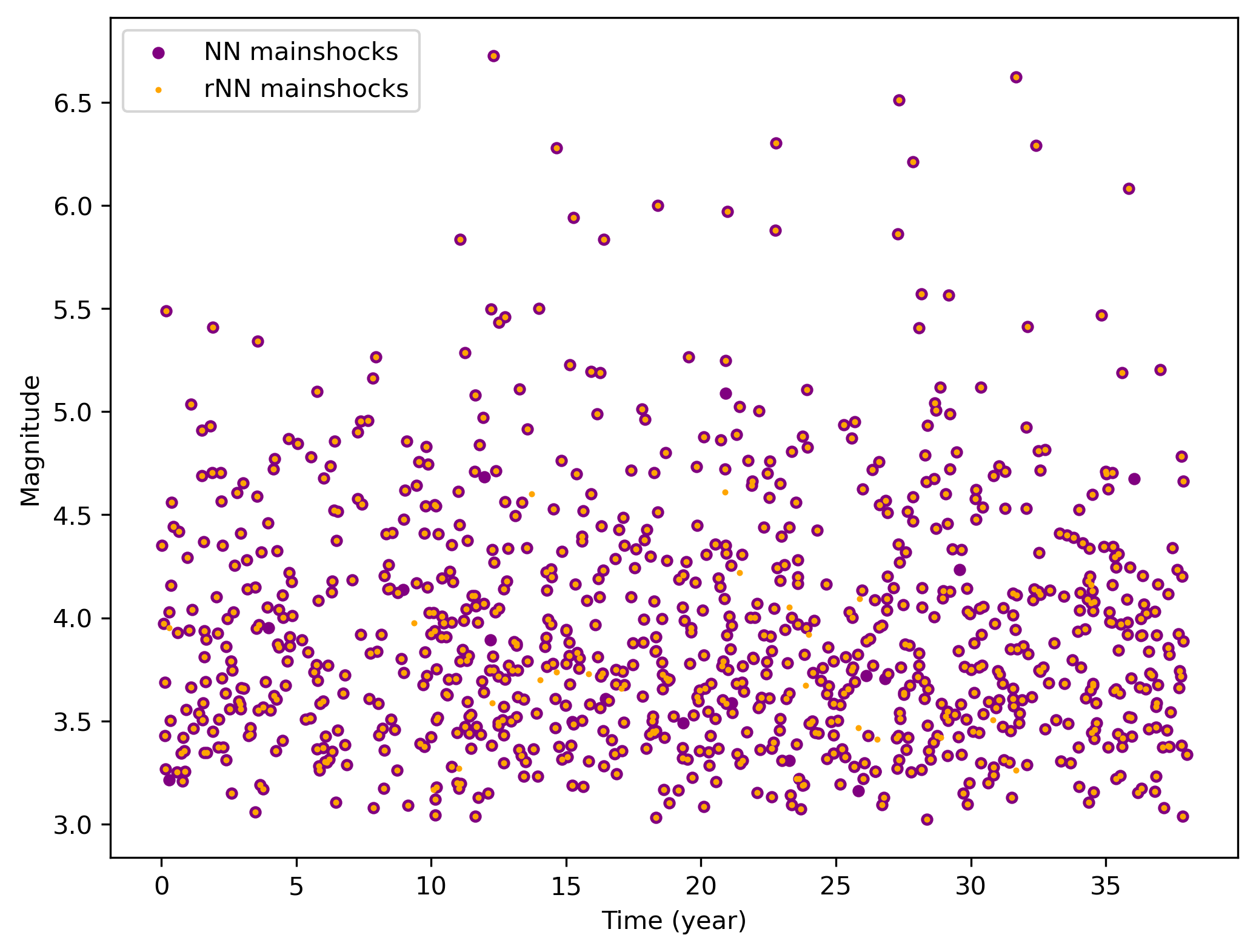}
\includegraphics[width=0.48\textwidth]{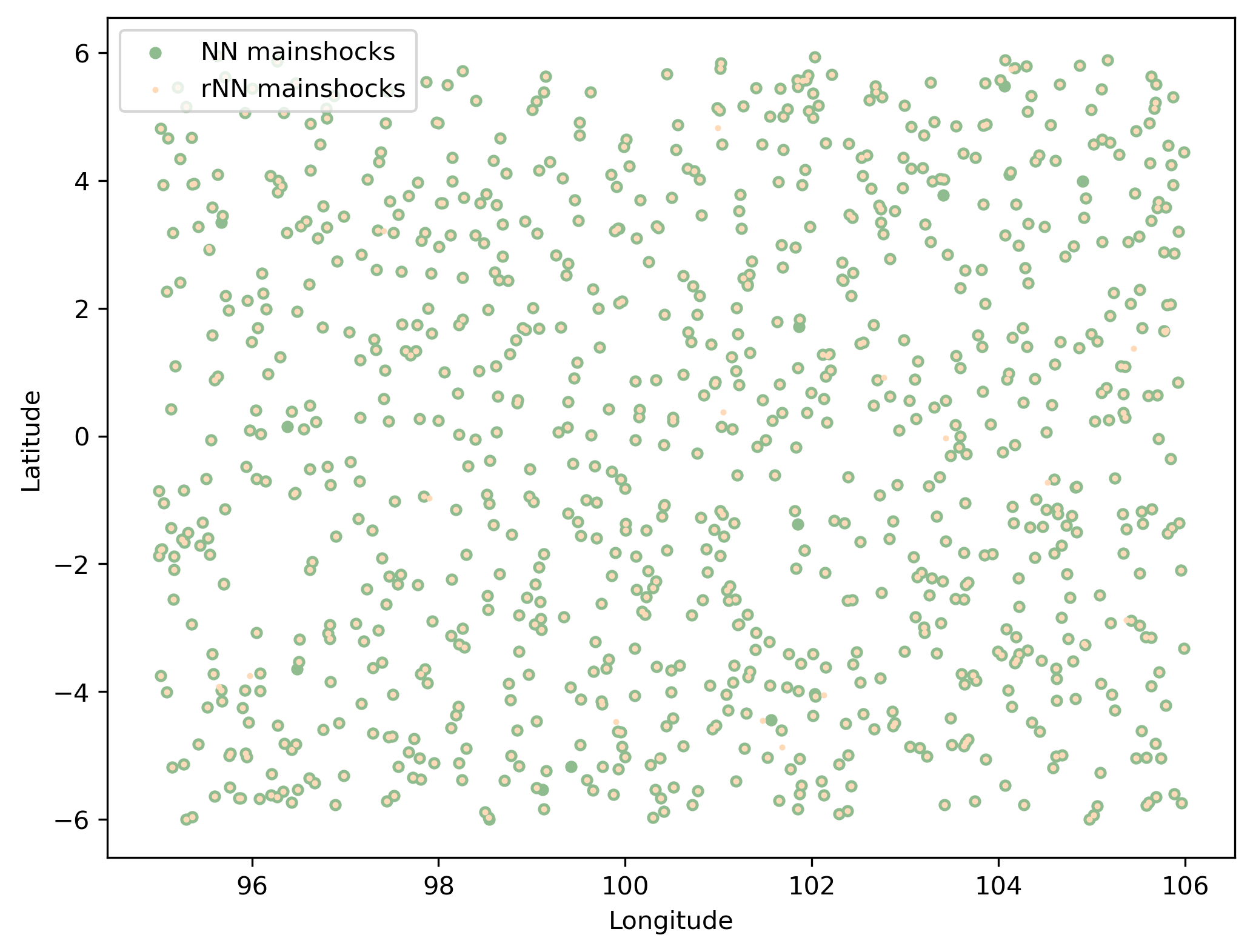}
\caption{
Comparison between the mainshock populations identified by the standard NN
and reverse nearest-neighbor (rNN) declustering procedures.
Left panel: magnitude-time distribution of NN and rNN mainshocks.
Right panel: spatial distribution of NN and rNN mainshocks.
The two methods identify nearly the same mainshock population. This indicates
that hybrid procedures using NN-defined mainshocks and adding rNN-identified
foreshocks are expected to give results very close to the full rNN analysis.
}
\label{fig:mainshock_comparison}
\end{figure}

We also tested the hybrid strategy suggested by comparing the mainshock population identified by NN and rNN. This test is important because a possible hybrid approach consists in defining mainshocks using the standard NN declustering and then adding the additional foreshocks identified by rNN for the same mainshocks. Figure~\ref{fig:mainshock_comparison} shows that the mainshocks identified by the two methods are almost identical, both in their time-magnitude distribution and in their spatial distribution.

This result is important because it demonstrates that the differences reported
later in the foreshock statistics are not caused by a fundamentally different
separation between clustered and independent seismicity. Instead, NN and rNN
mainly differ in how events are internally organized within clusters, namely in
the assignment of precursor-mainshock relationships. In other words, the main
effect of reversing the nearest-neighbor symmetry is not to redefine the global
cluster structure, but rather to modify the inferred genealogy of events inside
already clustered seismicity.\\

Having established that NN and rNN produce very similar
background-triggered partitions, we now investigate whether the two
methods differ in their reconstruction of foreshock populations.

\subsubsection{ETAS catalogues}

We first investigate the performance of NN and rNN declustering on synthetic catalogues generated using the ETAS model, corresponding to a purely bottom-up triggering regime. In these simulations, all foreshocks emerge exclusively from stochastic earthquake-to-earthquake triggering cascades, without any explicit top-down preparatory component. Consequently, ETAS catalogues provide a controlled benchmark for evaluating whether the declustering procedures are able to correctly reconstruct bottom-up foreshock organization. \\
For each pair of magnitude classes \((F,M)\), we compute the distribution of the number of foreshocks per catalogue using three different definitions: the true foreshock genealogy known from the ETAS simulation, the reconstruction obtained using standard nearest-neighbor (NN) declustering, and the reconstruction obtained using reverse nearest-neighbor (rNN) declustering.

\begin{figure}[htbp]
    \centering
    \includegraphics[width=\linewidth]{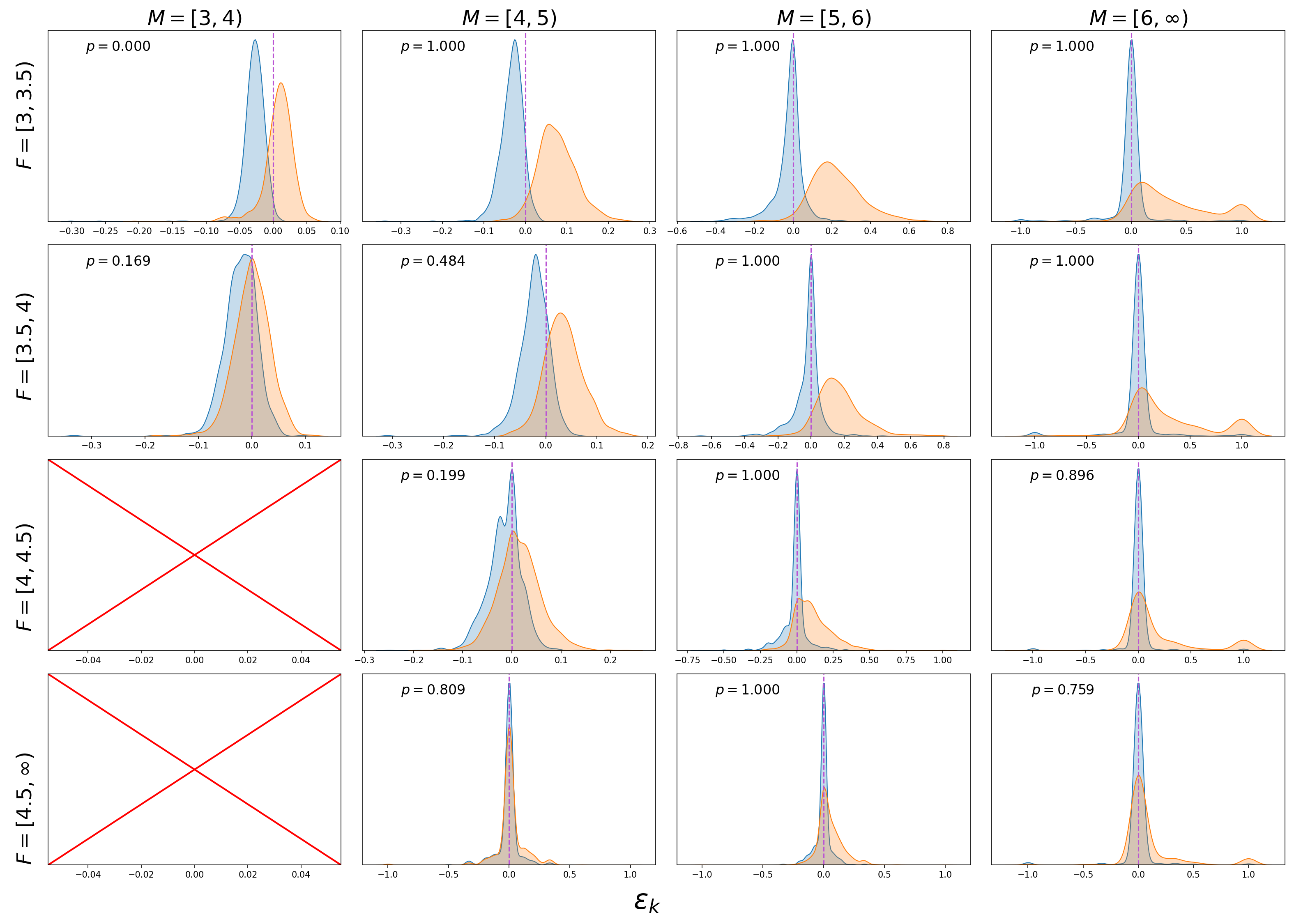}
    \caption{
    Distributions of the normalized foreshock-count residual \(\epsilon_k\) obtained from synthetic ETAS catalogues for different foreshock--mainshock magnitude classes. Rows correspond to foreshock magnitude ranges \(F\), while columns correspond to mainshock magnitude ranges \(M\). Blue curves correspond to the standard nearest-neighbor (NN) declustering method, whereas orange curves correspond to the reverse nearest-neighbor (rNN) method. The vertical value \(\epsilon_k=0\) represents perfect agreement with the true synthetic genealogy; negative values indicate underestimation of foreshock counts, while positive values indicate overestimation. The values reported in each panel give the bootstrap comparison between NN and rNN.
    }
    \label{fig:ETAS_Counting}
\end{figure}

Figure~\ref{fig:ETAS_Counting} shows the distributions of the normalized foreshock-count residuals obtained for the ETAS simulations. Rows correspond to different foreshock magnitude classes, while columns correspond to different mainshock magnitude classes. In each panel, the blue curve represents the residual distribution obtained with standard NN declustering, whereas the orange curve represents the residual distribution obtained with rNN declustering. Values close to zero indicate agreement with the true synthetic
genealogy.

Several important features emerge from Figure~\ref{fig:ETAS_Counting}. For small foreshock and mainshock magnitudes, the rNN residual distributions are generally closer to zero than the NN distributions. In these bins, the standard NN metric tends to underestimate the number of foreshocks, whereas rNN partially reduces this negative bias and more accurately reproduces the true synthetic genealogy.

As the foreshock and mainshock magnitudes increase, the differences between the two methods progressively decrease. For moderate and large magnitude classes, NN and rNN produce broadly comparable residual distributions, with neither method showing a clear and systematic advantage across all magnitude bins. In particular, the residual distributions become increasingly overlapping, indicating that both metrics reconstruct the true foreshock population with similar accuracy.

This behaviour is physically consistent with the structure of the ETAS model. Since ETAS is a purely bottom-up triggering process, the causal organization of seismicity is naturally forward in time. Nevertheless, for small earthquakes, the temporal asymmetry encoded in the NN metric appears to introduce a modest underestimation of foreshock activity. Reversing the declustering symmetry partially compensates for this effect, resulting in a more accurate reconstruction of the true genealogy in the low-magnitude regime.

To quantify the relative performance of the two declustering methods, we compare the reconstructed foreshock-count distributions with the true distribution obtained from the synthetic genealogy.

For each foreshock--mainshock magnitude bin $(i,j)$, the ensemble of synthetic catalogues provides the true distribution of foreshock counts,
\[
P_{\rm true}^{ij}(k),
\]
together with the distributions reconstructed using NN and rNN declustering,
\[
P_{\rm NN}^{ij}(k),
\qquad
P_{\rm rNN}^{ij}(k),
\]
where $k$ denotes the number of foreshocks observed in a given catalogue.

For each method $m\in\{\mathrm{NN},\mathrm{rNN}\}$, we compute the cross-entropy score

\[
\mathcal{L}_{m}^{ij}
=
\sum_k
P_{\rm true}^{ij}(k)
\log
\left[
\max
\left(
P_m^{ij}(k),
\varepsilon
\right)
\right],
\]

with $\varepsilon=10^{-9}$ introduced to avoid numerical divergences when an empirical probability is zero. Larger values of $\mathcal{L}_{m}^{ij}$ indicate better agreement with the true foreshock-count distribution.

We then define

\[
\Delta\mathcal{L}^{ij}
=
\mathcal{L}_{\rm rNN}^{ij}
-
\mathcal{L}_{\rm NN}^{ij}.
\]

Positive values indicate that rNN provides a more accurate reconstruction of the true foreshock-count distribution, whereas negative values favour NN.

To estimate statistical significance, we perform a bootstrap procedure based on the true foreshock-count distribution. For each bootstrap realization $b$, we generate a resampled estimate of the true distribution and recompute

\[
\Delta\mathcal{L}^{(b),ij}.
\]

The reported $p$-value is defined as

\[
p
=
\frac{
\#\left\{
b:
\Delta\mathcal{L}^{(b),ij}<0
\right\}
}{
B
},
\]

where $B=1000$ is the number of bootstrap realizations. Therefore, $p$ represents the bootstrap probability that NN performs better than rNN. The $p$-values are reported in Figure~\ref{fig:ETAS_Counting} in each panel.

\begin{figure}[htbp]
\centering
\includegraphics[width=\linewidth]{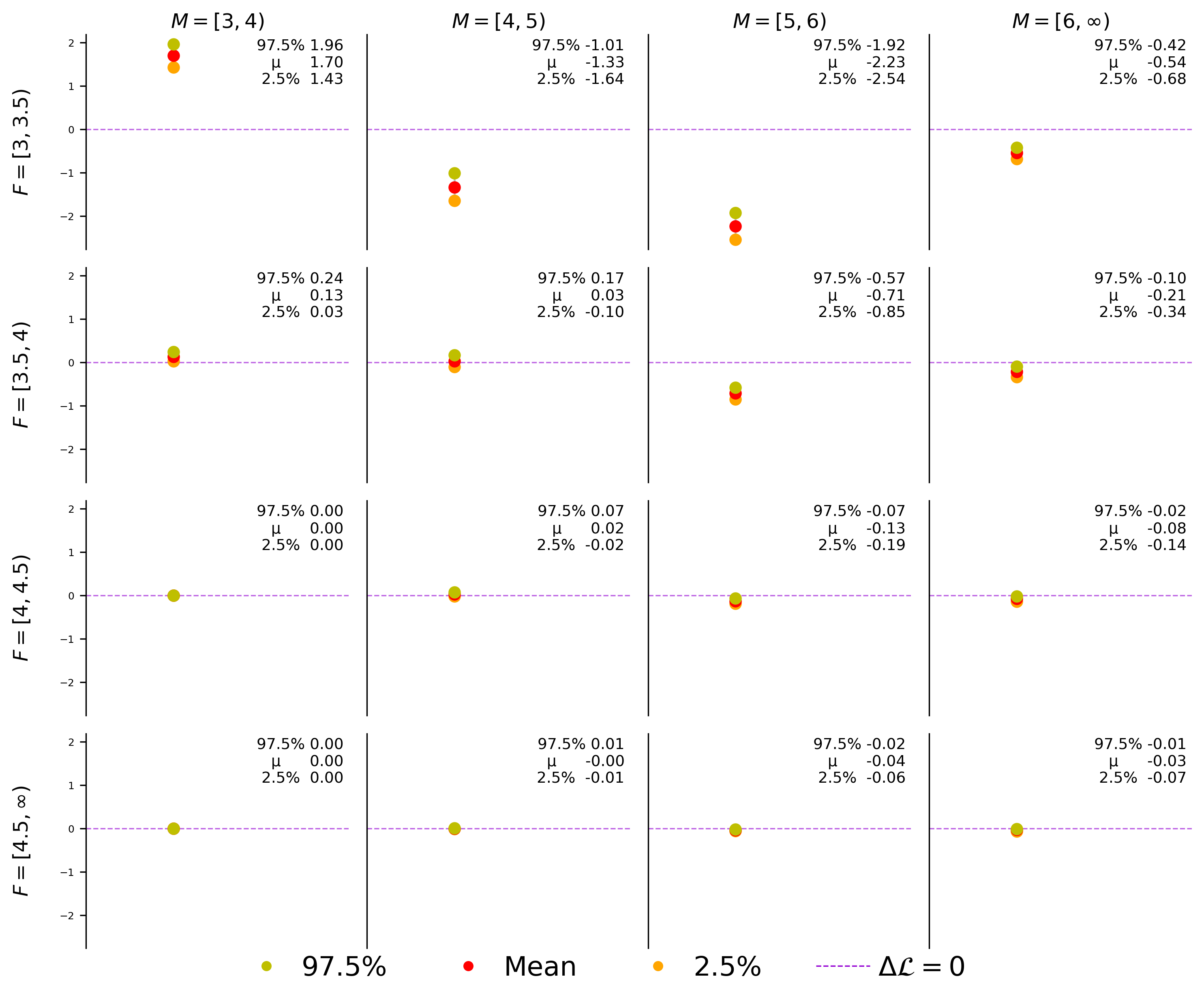}
\caption{
Bootstrap cross-entropy comparison between rNN and NN for synthetic ETAS catalogues. Each panel corresponds to a foreshock--mainshock magnitude bin. The red symbol indicates the mean value of \(\Delta\mathcal{L}^{ij}=\mathcal{L}_{\rm rNN}^{ij}-\mathcal{L}_{\rm NN}^{ij}\), while the orange and yellow symbols indicate the 2.5\% and 97.5\% bootstrap quantiles, respectively. The dashed horizontal line marks \(\Delta\mathcal{L}=0\). Positive values indicate that rNN provides a better reconstruction of the true foreshock-count distribution, whereas negative values favour NN.
}
\label{fig:ETAS_likelihood}
\end{figure}

Figure~\ref{fig:ETAS_likelihood} summarizes the same bootstrap cross-entropy comparison in terms of the mean value of
\(\Delta\mathcal{L}^{ij}\) and its 95\% confidence interval for each foreshock--mainshock magnitude bin. Consistently with Figure~\ref{fig:ETAS_Counting}, positive values are mainly concentrated in the low-magnitude part of the matrix, especially for small foreshocks and small-to-intermediate mainshocks. This confirms that rNN improves the reconstruction of the true foreshock-count distribution in these bins. For larger foreshock and mainshock magnitudes, \(\Delta\mathcal{L}^{ij}\) approaches zero or becomes slightly negative, indicating that NN and rNN perform similarly, or that NN is marginally favoured in some high-magnitude bins.

\subsubsection{ETAFS catalogues}

We now investigate the performance of NN and rNN declustering on synthetic catalogues generated using the ETAFS model, corresponding to a mixed top-down/bottom-up seismicity regime. In contrast to ETAS, the ETAFS model explicitly incorporates a preparatory component through the direct generation of foreshocks around future large events. Consequently, the synthetic catalogues contain two fundamentally different types of foreshocks: bottom-up foreshocks generated through ordinary triggering cascades and top-down foreshocks associated with the preparatory organization of future mainshocks.

As in the ETAS analysis, we divide mainshocks and foreshocks into discrete magnitude classes and compute, for each magnitude pair \((F,M)\), the distribution of the number of foreshocks reconstructed using NN and rNN declustering. Since the true genealogy of the synthetic catalogues is known, we compare the reconstructed foreshock counts directly with the total true foreshock population generated by the ETAFS model.

\begin{figure}[htbp]
    \centering
    \includegraphics[width=\linewidth]{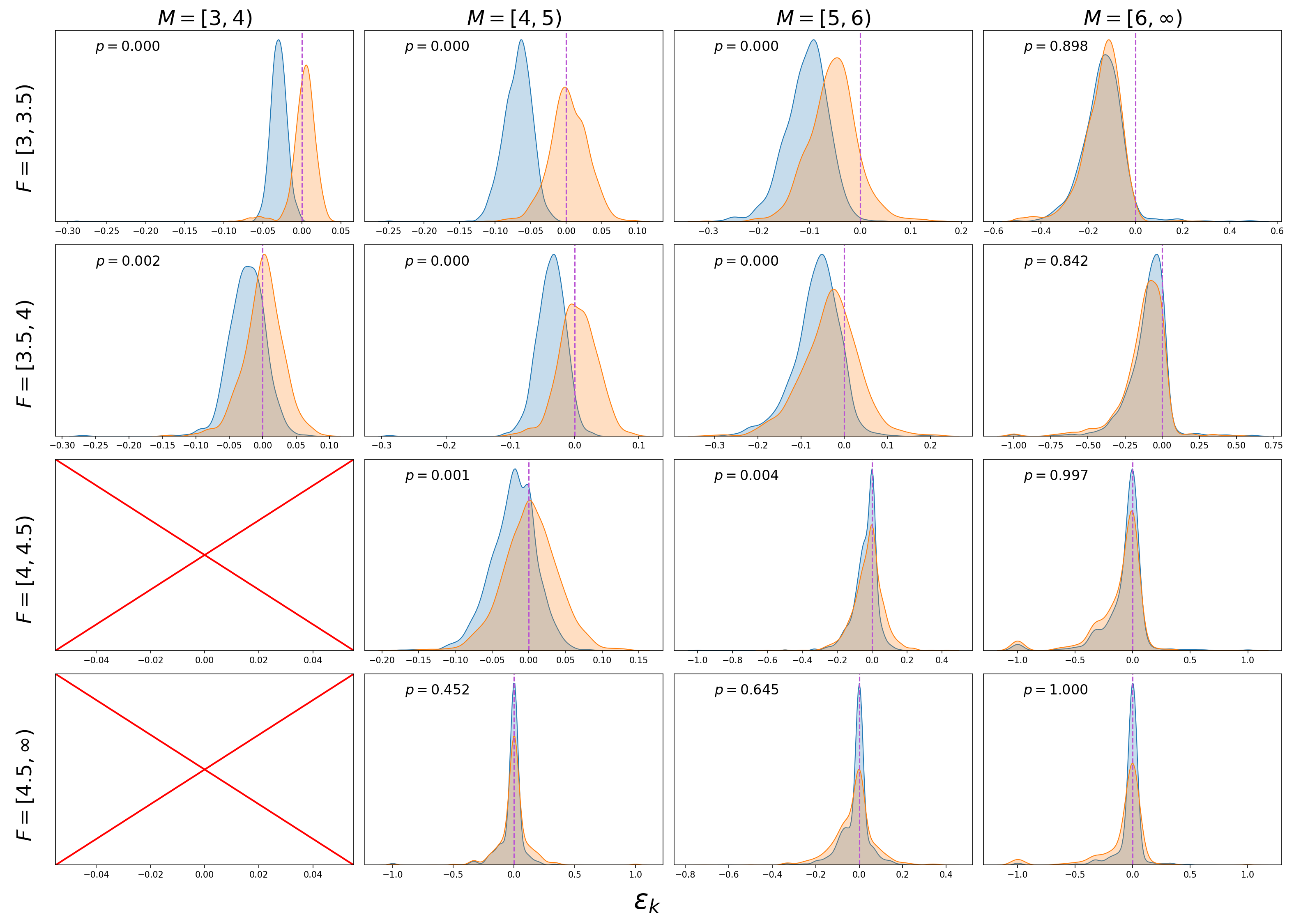}
    \caption{
    Distributions of the normalized foreshock-count residual \(\epsilon_k\) obtained from synthetic ETAFS catalogues for different foreshock--mainshock magnitude classes. Rows correspond to foreshock magnitude ranges \(F\), while columns correspond to mainshock magnitude ranges \(M\). Blue curves correspond to the standard nearest-neighbor (NN) declustering method, whereas orange curves correspond to the reverse nearest-neighbor (rNN) method. The value \(\epsilon_k=0\) represents perfect agreement with the true total foreshock population generated by the ETAFS model; negative values indicate underestimation and positive values indicate overestimation. The values reported in each panel give the bootstrap cross-entropy comparison between NN and rNN (corresponding $p$-value).
    }
    \label{fig:ETAFS_Counting}
\end{figure}

Figure~\ref{fig:ETAFS_Counting} shows the distributions of the normalized foreshock-count residuals obtained for the ETAFS simulations. Rows correspond to different foreshock magnitude classes, while columns correspond to different
mainshock magnitude classes. In each panel, the blue curve represents the residual distribution obtained with standard NN declustering, whereas the orange curve represents the residual distribution obtained with rNN declustering. Values close to zero indicate agreement with the true total foreshock population generated by the ETAFS model.

Several qualitative differences relative to the ETAS results immediately emerge. As in the ETAS catalogues, the largest differences between NN and rNN are concentrated in the low-magnitude portion of the foreshock--mainshock matrix.
For small foreshocks associated with small and intermediate mainshocks, the NN residual distributions are systematically shifted towards negative values, indicating an underestimation of foreshock occurrence. In these bins, rNN produces residual distributions that are closer to zero and therefore provides a more accurate reconstruction of the true foreshock population. As the foreshock and mainshock magnitudes increase, the differences between NN and rNN progressively decrease. For moderate and large magnitude classes, the two methods produce broadly similar residual distributions, indicating comparable reconstruction performance. Therefore, the ETAFS simulations do not suggest a uniform superiority of rNN across all magnitude ranges. Instead, they indicate that the advantage of rNN is concentrated in the same low-magnitude regime identified in the ETAS catalogues, although the improvement remains visible over a somewhat broader region of the foreshock--mainshock matrix. \\

\begin{figure}[htbp]
\centering
\includegraphics[width=\linewidth]{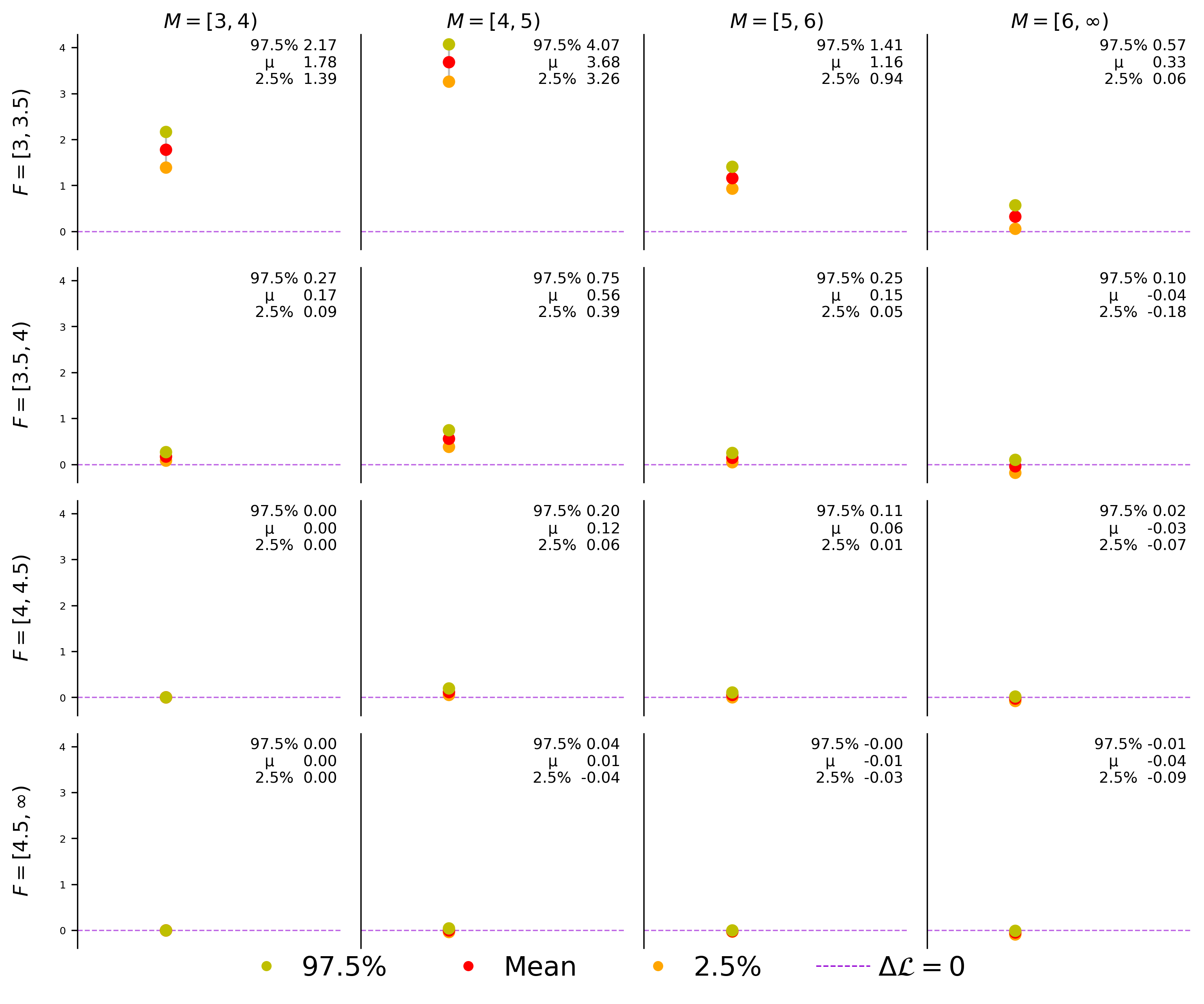}
\caption{
Bootstrap cross-entropy comparison between rNN and NN for synthetic ETAFS catalogues. Each panel corresponds to a foreshock--mainshock magnitude bin. The red symbol indicates the mean value of \(\Delta\mathcal{L}^{ij}=\mathcal{L}_{\rm rNN}^{ij}-\mathcal{L}_{\rm NN}^{ij}\), while the orange and yellow symbols indicate the 2.5\% and 97.5\% bootstrap quantiles, respectively. The dashed horizontal line marks \(\Delta\mathcal{L}=0\). Positive values indicate that rNN provides a better reconstruction of the true foreshock-count distribution, whereas negative values favour NN.
}
\label{fig:ETAFS_likelihood}
\end{figure}

The bootstrap cross-entropy comparison shown in Figure~\ref{fig:ETAFS_likelihood} confirms this interpretation. As for ETAS, the strongest positive values of \(\Delta\mathcal{L}^{ij}\) occur in the low-magnitude part of the matrix, indicating that rNN better reproduces the true ETAFS foreshock-count distribution in these bins. For larger magnitudes, the confidence intervals move closer to zero and the advantage of rNN becomes weaker or disappears. Thus, the ETAFS results do not indicate a uniform improvement of rNN over all magnitude classes, but rather a magnitude-dependent improvement concentrated mainly at low foreshock and mainshock magnitudes.

We have also verified that the same small-magnitude bias is observed when considering simulated catalogues with minimum magnitude equal to 2.

\subsection{Application to Southern California seismicity}

We finally apply the proposed declustering framework to the Southern California seismic catalogue, considering only events above the completeness magnitude threshold $m_c=3$. This conservative threshold was selected following the methodology proposed by \citeA{PZ23}, in order to minimize possible biases associated with short-term incompleteness and spatial heterogeneity of the catalogue.

\begin{figure*}
\centering
\includegraphics[width=\textwidth]{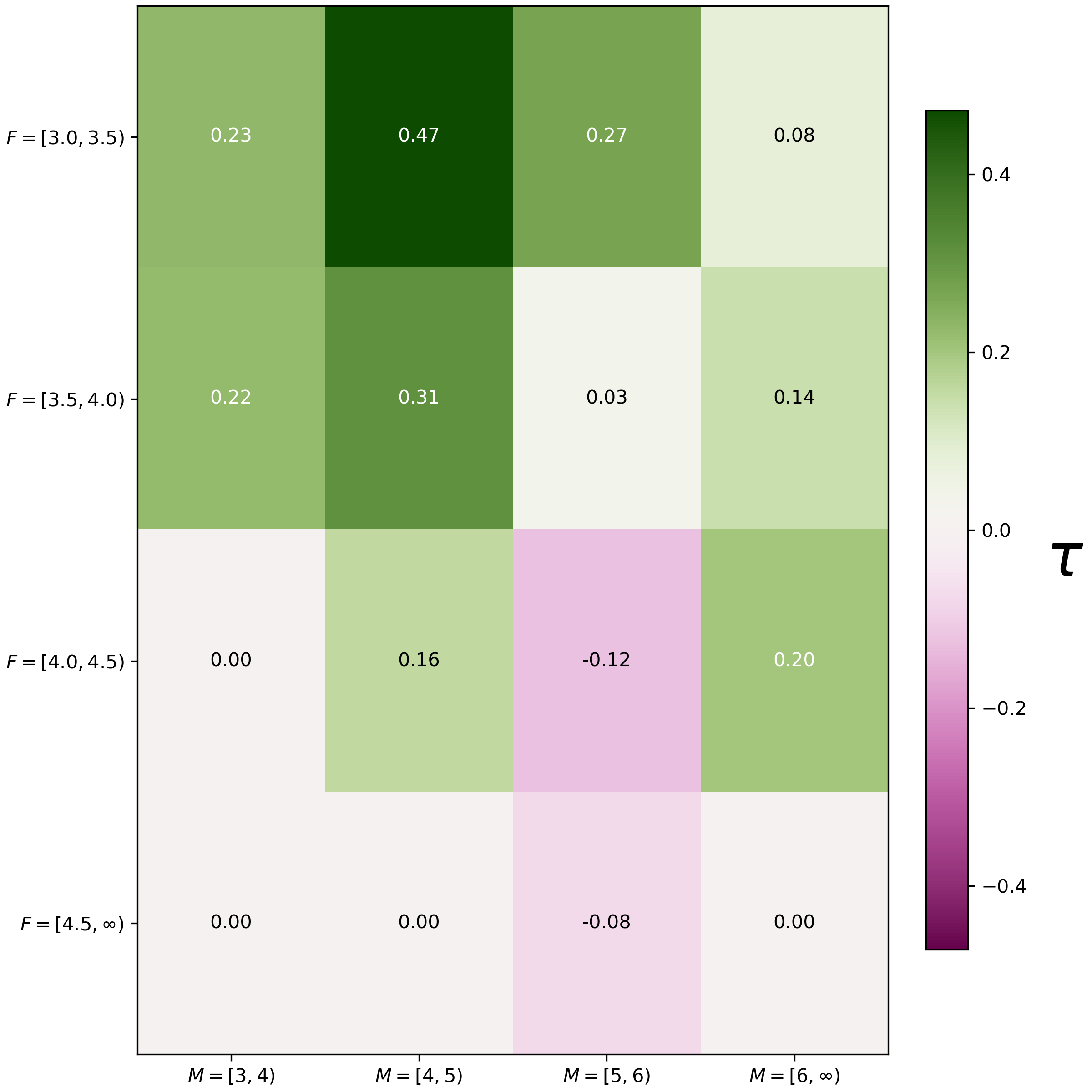}
\caption{Normalized difference between the number of foreshocks identified by the reverse nearest-neighbour (rNN) and the standard nearest-neighbour (NN) declustering methods in the Southern California catalogue. Each matrix element reports $\tau=\frac{N_{\rm rNN}-N_{\rm NN}}{N_{\rm rNN}+N_{\rm NN}}$,
where $N_{\rm rNN}$ and $N_{\rm NN}$ denote the numbers of foreshocks identified by the two methods within the corresponding foreshock ($F$) and mainshock ($M$) magnitude classes. Positive values indicate that rNN identifies more foreshocks than NN, whereas negative values indicate the opposite. The largest positive values are concentrated in the low- and intermediate-magnitude regime, indicating that the reverse declustering symmetry preferentially increases foreshock detection in these magnitude classes.}
\label{fig:real_catalogue_matrix}
\end{figure*}

To quantify the differences between the two declustering schemes in the real catalogue, we compute the normalized difference
\begin{equation}
    \tau=
\frac{N_{\rm rNN}-N_{\rm NN}}{N_{\rm rNN}+N_{\rm NN}},
\end{equation}
for each foreshock--mainshock magnitude class. This quantity ranges from $-1$ to $1$ and provides a scale-independent measure of the relative excess of foreshocks identified by one method with respect to the other. Values close to zero indicate that the two methods identify similar numbers of foreshocks, whereas positive (negative) values indicate that rNN (NN) identifies more foreshocks. Figure~\ref{fig:real_catalogue_matrix} shows the normalized difference $\tau$ between the foreshock populations identified by rNN and NN for each combination of foreshock and mainshock magnitude classes.

A clear systematic pattern emerges. Positive values dominate most of the matrix, demonstrating that rNN generally identifies more foreshocks than the standard NN method. The largest relative gains occur for small and intermediate foreshock magnitudes associated with intermediate mainshock magnitudes. In these bins, rNN identifies almost 30-40\% more foreshocks than NN. By contrast, for the largest magnitude classes the two methods become progressively more similar and $\tau$ approaches zero.

This behaviour is qualitatively consistent with the synthetic ETAS and ETAFS experiments. In both synthetic models, the largest improvements obtained with rNN were concentrated in the low- and intermediate-magnitude portion of the foreshock--mainshock matrix. The Southern California catalogue exhibits the same overall tendency, suggesting that the declustering symmetry primarily affects the reconstruction of relatively small foreshock populations, while the two methods converge for larger earthquakes.

From a physical perspective, this behaviour reflects the different temporal symmetries underlying the two declustering metrics. In the standard NN metric, the correlation distance is controlled by the magnitude of the earlier event through the factor $10^{-b m_i}$, whereas in rNN it is controlled by the magnitude of the later event through the factor $10^{-b m_j}$. This behaviour follows naturally from the different temporal symmetries of the two metrics. Because the rNN distance is scaled by the magnitude of the future event rather than by that of the preceding event, large future earthquakes are associated with a larger effective correlation domain when searching for possible foreshocks. Consequently, rNN is more likely to associate preceding earthquakes with the same future mainshock, increasing the inferred foreshock population in the corresponding magnitude bins.

The predominance of positive $\tau$ values therefore suggests that standard nearest-neighbour declustering systematically underestimates foreshock occurrence over a substantial portion of the magnitude space. Although the true genealogy of the Southern California catalogue is unknown, the synthetic experiments indicate that this behaviour is an intrinsic consequence of the temporal asymmetry of the declustering metric rather than necessarily evidence for a preparatory physical mechanism. The real catalogue therefore confirms that the inferred abundance of foreshocks depends appreciably on the adopted declustering symmetry.

These results have two important implications. First, they suggest that conclusions regarding the rarity or abundance of foreshocks may depend significantly on the adopted declustering metric. Second, they indicate that declustering procedures calibrated on standard ETAS assumptions may not be fully adequate when the seismicity contains precursor-like structures or correlations not captured by purely bottom-up triggering models.

To complement the foreshock-count comparison, we also computed the probability that a mainshock is associated with at least one foreshock. For each mainshock magnitude class, this probability is defined as the fraction of mainshocks for which at least one foreshock is identified by the corresponding declustering method.

\begin{figure}[htbp]
\centering
\includegraphics[width=0.90\textwidth]{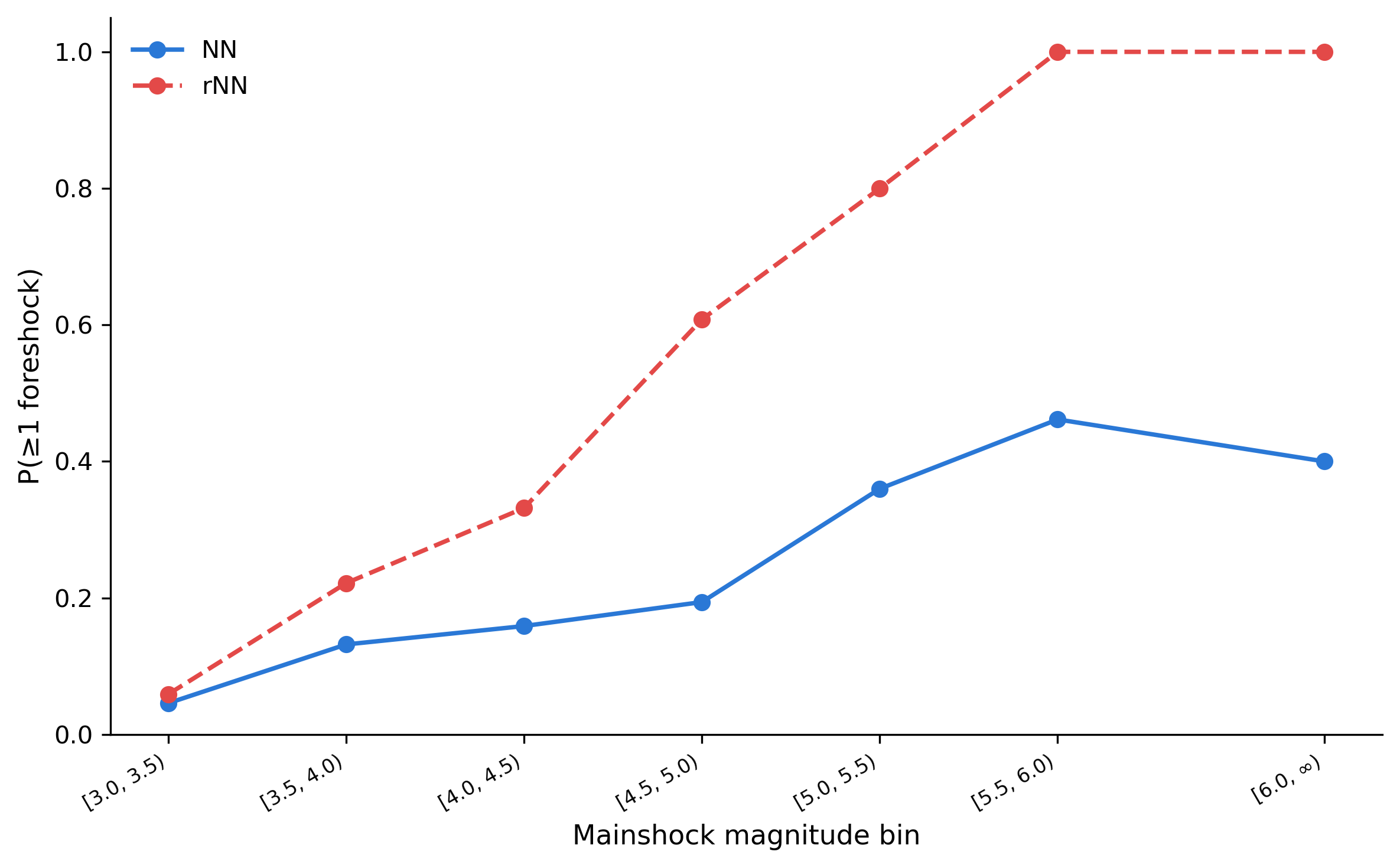}
\caption{
Probability that a mainshock is associated with at least one identified foreshock as a function of mainshock magnitude. Blue symbols correspond to the standard nearest-neighbour (NN) declustering method, while red symbols correspond to the reverse nearest-neighbour (rNN) method. For each magnitude bin, the probability is computed as the fraction of mainshocks with at least one foreshock identified by the corresponding declustering procedure. The rNN method yields systematically higher foreshock probabilities than NN, indicating that reversing the declustering symmetry increases not only the number of identified foreshocks but also the fraction of mainshocks preceded by foreshock activity.
}
\label{fig:foreshock_probability}
\end{figure}

Figure~\ref{fig:foreshock_probability} shows that rNN yields a systematically higher foreshock probability than NN across all mainshock magnitude classes. This indicates that the larger number of foreshocks identified by rNN is not only due to a few highly active sequences, but also corresponds to a larger fraction of mainshocks being classified as preceded by foreshock activity. This result further supports the conclusion that standard NN declustering may underestimate foreshock occurrence in the Southern California catalogue.

\section{Conclusions}

In this work, we investigated whether the number of foreshocks inferred from seismic catalogues depends on the declustering metric used to identify earthquake clusters. Motivated by the long-standing debate between bottom-up triggering and top-down preparatory interpretations of foreshocks \cite{mignan2014debate}, we compared the classical nearest-neighbor declustering metric \cite{BP04,ZGKW08,ZB13} with a reverse nearest-neighbor metric, denoted rNN. The proposed rNN metric is obtained by assigning the Gutenberg-Richter magnitude factor to the later event rather than to the earlier one. Equivalently, it corresponds to applying the standard nearest-neighbor construction to a time-reversed catalogue.

Using synthetic ETAS catalogues \cite{ogata1988,ogata1998}, we showed that NN and rNN produce broadly comparable results over much of the foreshock--mainshock magnitude matrix. The main differences emerge in the low-magnitude regime, where rNN more accurately reconstructs the true foreshock population and reduces the negative bias observed with NN declustering.

A key result of this study is that the largest differences between NN and rNN are concentrated in the low-magnitude portion of the foreshock--mainshock matrix. In both ETAS and ETAFS catalogues, rNN systematically reduces the underestimation of foreshocks associated with small foreshocks and small mainshocks. As event magnitudes increase, the differences between the two methods progressively decrease and the inferred foreshock populations become increasingly similar.
This observation suggests that the main limitation of the standard nearest-neighbour metric is not necessarily related to the physical origin of foreshocks, but rather to the reconstruction of low-magnitude precursor activity. The benefit of reversing the declustering symmetry therefore appears in both purely bottom-up and mixed top-down/bottom-up seismicity models.

The application to Southern California further supports the practical relevance of this conclusion. Since the true genealogy is unknown in real catalogues, the comparison between NN and rNN cannot by itself determine the physical origin of foreshocks. Nevertheless, the fact that rNN identifies systematically more foreshocks than NN in several magnitude classes, and assigns at least one foreshock to a larger fraction of mainshocks, indicates that the inferred foreshock population is not invariant under the choice of declustering metric. This suggests that standard declustering may systematically underestimate foreshock activity in the low-magnitude regime, where earthquake catalogues contain the largest number of events and therefore much of the available statistical information.

Overall, our results show that foreshock detectability depends on the declustering symmetry and on the magnitude regime considered. Our findings are consistent with the conclusions of \citeA{christophersen2008foreshock}, who argued that foreshock occurrence rates are highly sensitive to the adopted classification procedure. While their study focused on the definitions of foreshocks and aftershocks within earthquake clusters, our results indicate that the declustering metric itself introduces an additional source of variability. Consequently, differences in reported foreshock rates between studies may reflect not only physical differences between seismic sequences but also differences in the methodology used to reconstruct earthquake genealogies. The number of foreshocks is not only a property of the seismic catalogue, but also of the temporal symmetry implicitly assumed by the declustering procedure. This has important implications for earthquake forecasting and for statistical tests of the foreshock hypothesis. In particular, rejecting or supporting a top-down preparatory interpretation using declustered catalogues requires caution, because the declustering method itself may suppress or enhance the foreshock population being tested.

Future work should extend this comparison to additional tectonic regions, explore adaptive combinations of NN and rNN metrics, and test whether the discrepancy between forward and reverse declustering can be used as an empirical diagnostic of preparatory seismic organization. Such developments may help transform declustering from a purely classificatory tool into a quantitative probe of the physical mechanisms underlying earthquake clustering.

\section*{Open Research Section}
All source codes to reproduce the analyses and figures presented in this study are publicly available in the Zenodo repository: \url{https://doi.org/10.5281/zenodo.21260067}. The repository contains all resources required to reproduce the synthetic simulations, declustering analyses, and the results reported in this paper.

\section*{Conflict of Interest declaration}
The authors declare there are no conflicts of interest for this manuscript.

\acknowledgments
E.L. acknowledges support from the MIUR PRIN 2022 PNRR P202247YKL. G.P. acknowledges support from the DyDéCo project from the CPJ Chair at ENS de Lyon and CNRS.

\bibliography{Reference.bib}

\end{document}